\documentclass[mathleft
]{an}
\usepackage{graphicx}
\usepackage{rotating}
\usepackage{lscape}
\usepackage{longtable}
\usepackage{times}
\usepackage{fge}
\begin{document}

\Pagespan{789}{}
\Yearpublication{2011}%
\Yearsubmission{2010}%
\Month{11}%
\Volume{999}%
\Issue{88}%

\sloppy

\title{An Arabic report about supernova SN 1006 by Ibn S{\={\i}}n\={a} (Avicenna)}

\author{R. Neuh\"auser\inst{1} \thanks{Corresponding author: \email{rne@astro.uni-jena.de}}
\and
C. Ehrig-Eggert\inst{2} 
\and
P. Kunitzsch\inst{3}
}

\titlerunning{SN 1006 by Ibn S{\={\i}}n\={a}}
\authorrunning{Neuh\"auser et al.}

\institute{
Astrophysikalisches Institut und Universit\"ats-Sternwarte, FSU Jena,
Schillerg\"a\ss chen 2-3, 07745 Jena, Germany (e-mail: rne@astro.uni-jena.de)
\and
Institut f\"ur Geschichte der Arabisch-Islamischen Wissenschaften, 
Westendstra\ss e 89, 60325 Frankfurt a.M., Germany (retired)
\and
LMU Munich, Germany (retired); home: Davidstrasse 17, 81927 M\"unchen, Germany
}

\received{2016 Feb 17}
\accepted{2016 Apr 8}
\publonline{ }


\abstract{We present here an Arabic report about supernova 1006 (SN 1006)
written by the famous Persian scholar Ibn S{\={\i}}n\={a} (Lat. Avicenna, AD 980-1037),
which was not discussed in astronomical literature before.
The short observational report about a new star is part of Ibn S{\={\i}}n\={a}'s book called {\em al-Shif\={a}'},
a work about philosophy including physics, astronomy, and meteorology.
We present the Arabic text and our English translation.
After a detailed discussion of the dating of the observation,
we show that the text specifies that the transient celestial object was
stationary and/or tail-less ({\em a star among the stars}),
that it {\em remained for close to three months getting fainter and fainter until it disappeared},
that it {\em threw out sparks}, i.e. it was scintillating and very bright, 
and that the colour changed with time.
The information content is consistent with the other Arabic and non-Arabic reports about SN 1006.
Hence, it is quite clear that Ibn S{\={\i}}n\={a} refers to SN 1006 in his report, 
given as an example for transient celestial objects in a discussion of Aristotle's {\em Meteorology}.
Given the wording and the description, e.g. for the colour evolution, 
this report is independent from other reports known so far.
}

\maketitle

\section{Introduction: Supernova 1006}

Historic observations of supernovae (SN) are important to understand SNe, neutron stars, and SN remnants (SNR):
Historic reports can deliver the date of the observation (hence, the age of the SNR and, if existing, 
of the neutron star)
together with a light curve (hence, possibly the SN type), sometimes the colour and its evolution, 
and the position of the SN,
which is needed to identify the SNR and, if existing, the neutron star and/or pulsar wind nebula.
Such historic observations have been used very successfully for SNe
1006 (from Eastern Asia, Arabia, and Europe),
1054 (from Eastern Asia and Arabia),
1181 (only from Eastern Asia),
and SNe 1572 and 1604 (from Eastern Asia and Europe),
plus a few more SNe from the 1st millennium AD (see Stephenson \& Green 2002, henceforth SG02, and references therein).
While the Arabic report about SN 1054 merely confirms a bright new star in Gemini/Taurus around AD 1054,
the Arabic reports about SN 1006 present a lot of detailed information 
(Goldstein 1965; Cook 1999; SG02; Rada \& Neuh\"auser 2015).

According to historic observations and follow-up observations, 
SN 1006 and its SNR G327.6+14.6 have a distance 
of $2.18 \pm 0.08$ kpc with very small extinction (Winkler et al. 2003);
several arguments speak for a SN type Ia explosion (see Schaefer 1996);
for a SN type Ia, the peak apparent brightness would then be $-7.5 \pm 0.4$ mag (Winkler et al. 2003).

SN 1006 was observed by the Yemeni observer(s) around Apr 17/18 (Rada \& Neuh\"auser 2015),
by $^{c}$Al\={\i} ibn Ri\d{d}w\={a}n since Apr 30 (Goldstein 1965; SG02),
and in China and Japan since the end of April or early May (SG02).
The positional information by $^{c}$Al\={\i} ibn Ri\d{d}w\={a}n
(ecliptic longitude range) led to the identification of the SNR
(together with the right ascension range from the Chinese and the
declination limit from St. Gallen), see Stephenson et al. (1977) and SG02.
Several Arabic observers noted stationarity.
The report of Ibn Ab\={\i} Zar$^{c}$ 
(died in or after AD 1326) from a Moroccon source about SN 1006 (Goldstein 1965)
-- based on the edition of the Arabic and Latin text by Tornberg (1843) --
is the only source possibly 
mentioning a day-time observation: {\em Its appearance was before sunset ...} (SG02).

The following Arabic terms were used for historic observations of SNe:
\begin{itemize}
\item {\em kawkab}, which means {\em star} or {\em planet}, or more generally {\em celestial object},
used e.g. for SN 1006 by Ibn al-Jawz{\={\i}}, Ibn al-Ath\={\i}r, and Ibn Ab\={\i} Zar$^{c}$ (Goldstein 1965),
\item {\em najm}, which means just {\em star}, e.g. SN 1006 by al-Yam\={a}n{\={\i}} and Ibn al-Dayba$^{c}$ (Rada \& Neuh\"auser 2015),
\item {\em nayzak}, which can mean a {\em comet} or {\em new star}, e.g. SN 1006 by $^{c}$Al\={\i} ibn Ri\d{d}w\={a}n 
and Ibn Ab\={\i} Zar$^{c}$ (Goldstein 1965),
but also something like {\em spectacle} or {\em transient celestial event},
\item {\em athar}, which means {\em trace}, but which was also used for SN 1006 and SN 1054, and
\item {\em kawkab athar\={\i}} as {\em spectacular star} for SN 1054 by Ibn Ab\={\i} U\d{s}aybi$^{c}$a (Brecher et al. 1978).
\end{itemize}
If the observed object is classified just as some kind of {\em star},
but the duration of appearance is limited (to e.g. a few months), 
the object can be identified as transient. 
The class of {\em transient celestial objects} is often characterized by further details,
whether e.g. star-like, stationary, and/or with or without tail,
which then classifies it as, e.g., nova, SN, or comet. 
See Kunitzsch (1995) for a review of the Arabic words used for stars and transient celestial objects.

In their text book on historic SNe, Stephenson \& Green (2002) write in the chapter on {\em Future Prospects}: 
\begin{quotation}
In our view Arab writings have real potential as sources of further records of this SN [1006] -- and possibly
of that of AD 1054.
\end{quotation}
Indeed, we present here such a new record of SN 1006.

We present the Arabic text and our English translation in Sect. 2.
Then, in Sect. 3, we date the observation and interpret the text. We summarize our findings in Sect. 4.

\section{Ibn S{\={\i}}n\={a} and his report about SN 1006}

Ab\={u} $^{c}$Al{\={\i}} al-\d{H}usain b. $^{c}$Abdall\={a}h b. S{\={\i}}n\={a} 
(short Ibn S{\={\i}}n\={a}, Lat. Avicenna) was a Persian polymath\footnote{Even though he 
was a {\em Persian} scholar,
we think that it is correct to speak here of an {\em Arabic} report about SN 1006, 
because the transmitted text itself is written in Arabic.} 
and lived from AD 980-1037;
he wrote books about theology, medicine, and natural sciences including astronomy;
Ibn S{\={\i}}n\={a} follows in most topics Aristotle and Ptolemy,
but also tried to improve on the quality and quantity of celestial measurements (see, e.g., Sezgin 1978).
He invented the Jacob's staff or cross staff (Lat.: Baculus Jacobi) for precise altitude measurements 
(Wiedemann 1927), later replaced by the sextant.
In his works on the {\em Almagest} and in {\em al-Shif\={a}'}, Ibn S{\={\i}}n\={a} describes some of his own observations,
including what he interpreted as Venus transit,\footnote{In book IX of his {\em Almagest} and
very similar in the astronomy chapter of {\em al-Shif\={a}'}: {\em I say (that) I saw (myself) Venus
as/like a black dot (spot) on the surface/disc of the sun}, 
given without date (Goldstein 1969; Kapoor 2013); this is confirmed by two later
authors, namely by Na\d{s}{\={\i}}r al-D\={\i}n al-\d{T}\={u}s{\={\i}} 
({\em al Shaikh Ab\={u} $^{c}$Al{\={\i}} b. S{\={\i}}n\={a} mentions in his books that he
had seen Venus as a spot on the surface of the sun}) and by Yahuda b. Solomon Kohen ({\em Avicenna saw Venus
appearing like a spot in the midst of the sun}) (Goldstein 1969).} which was either a sunspot
or the Venus transit of AD 1032 May 24 (Goldstein 1969; Kapoor 2013).

Ibn S{\={\i}}n\={a}'s encyclopaedic book 
entitled {\em Kit\={a}b al-Shif\={a}'} (Book of Healing) is his major work on philosophy,
written from about AD 1013 to 1023; a nearly complete manuscript is located in the Bodleian Library, UK;
a critical edition of the Arabic text has been published by Madk\={u}r et al. (1965),
which we have used for our work (see Fig. 1).
In that work, Ibn S{\={\i}}n\={a} discussed Aristotelian philosophy including natural sciences.
During the discussion of Aristotle's {\em Meteorology} about transient celestial phenomena 
in the fifth volume, he mentioned a new star seen in 397h (AD 1006-1007).

We would like to remark how this short text (Fig. 1) about what is most certainly SN 1006 was found:
In his review of Sezgin (1979), Goldstein (1982) reported that he (Goldstein) was told by
A.I. Sabra that there is a mention of SN 1006 in Ibn S{\={\i}}n\={a}'s {\em Kit\={a}b al-Shif\={a}'}: 
\begin{quotation}
Professor A.I. Sabra informs me that a passage in Avicenna's {\em Meteorology} also mentions the supernova of 1006.
\end{quotation}

We present here the Arabic text from the edition of Madk\={u}r et al. (1965),
page 73, lines 12 to 17 (see Fig. 1):
\begin{quotation}
(line 12) {\em fa-ya$^{c}$ri\d{d}u li-dh\={a}lika an yabq\={a} iltih\={a}buh\={a} wa-ishti$^{c}$\={a}luh\={a}} \\
(line 13) {\em muddatan \d{t}aw{\={\i}}la imm\={a} $^{c}$al\={a} \d{s}\={u}rat dhu'\={a}ba aw dhanab, wa-aktharuhu sham\={a}l{\={\i}}
wa-qad yak\={u}nu jan\={u}b{\={\i}}yan, wa-imm\={a} $^{c}$al\={a}} \\
(line 14) {\em \d{s}\={u}rat kawkab min al-kaw\={a}kib, ka-lladh{\={\i}} \d{z}ahara f{\={\i}} sanat sab$^{c}$ wa-tis$^{c}${\={\i}}n
wa-thal\={a}th-mi'a li-l-hijra,} \\
(line 15) {\em fa-baqiya qar{\={\i}}ban min thal\={a}that ashhur yal\d{t}ufu wa-yal\d{t}ufu \d{h}att\={a} i\d{d}ma\d{h}alla,
wa-k\={a}na f{\={\i}} ibtid\={a}'ihi il\={a} l-saw\={a}d} \\
(line 16) {\em wa-l-khu\d{d}ra, thumma ja$^{c}$ala kull waqt yarm{\={\i}} bi-l-sharar wa-yazd\={a}du bay\={a}\d{d}an wa-yal\d{t}ufu \d{h}att\={a}
i\d{d}ma\d{h}alla, wa-qad} \\
(line 17) {\em yak\={u}nu $^{c}$al\={a} \d{s}\={u}rat li\d{h}ya, aw \d{s}\={u}rat \d{h}ayaw\={a}n lahu qur\={u}n, wa-$^{c}$al\={a} s\={a}'ir al-\d{s}uwar.}
\end{quotation}

Our English translation is as follows (words in round brackets are missing in one or 
some manuscripts, square brackets are our additions):
\begin{quotation}
It therefore happens that the burning and flaming stays for a (long) while,
either in form of a lock of hair or with a tail [i.e. in form of a comet],
mostly in the north, but sometimes also in the south,
or in form of a star among the stars [{\em kawkab min al-kaw\={a}kib}] -- 
like the one which appeared in the year 397(h).
It remained for close to three months [{\em qar{\={\i}}ban min thal\={a}that ashhur}] 
getting fainter and fainter until it disappeared;
at the beginning it was towards a darkness and greenness,
then it began to throw out sparks [{\em yarm{\={\i}} bi-l-sharar}] all the time, 
and then it became more and more whitish
and then became fainter and disappeared.
It can also have the form of a beard or of an animal with horns or of other figures.
\end{quotation}

\begin{figure*}
\begin{center}
\includegraphics[width=0.9\textwidth]{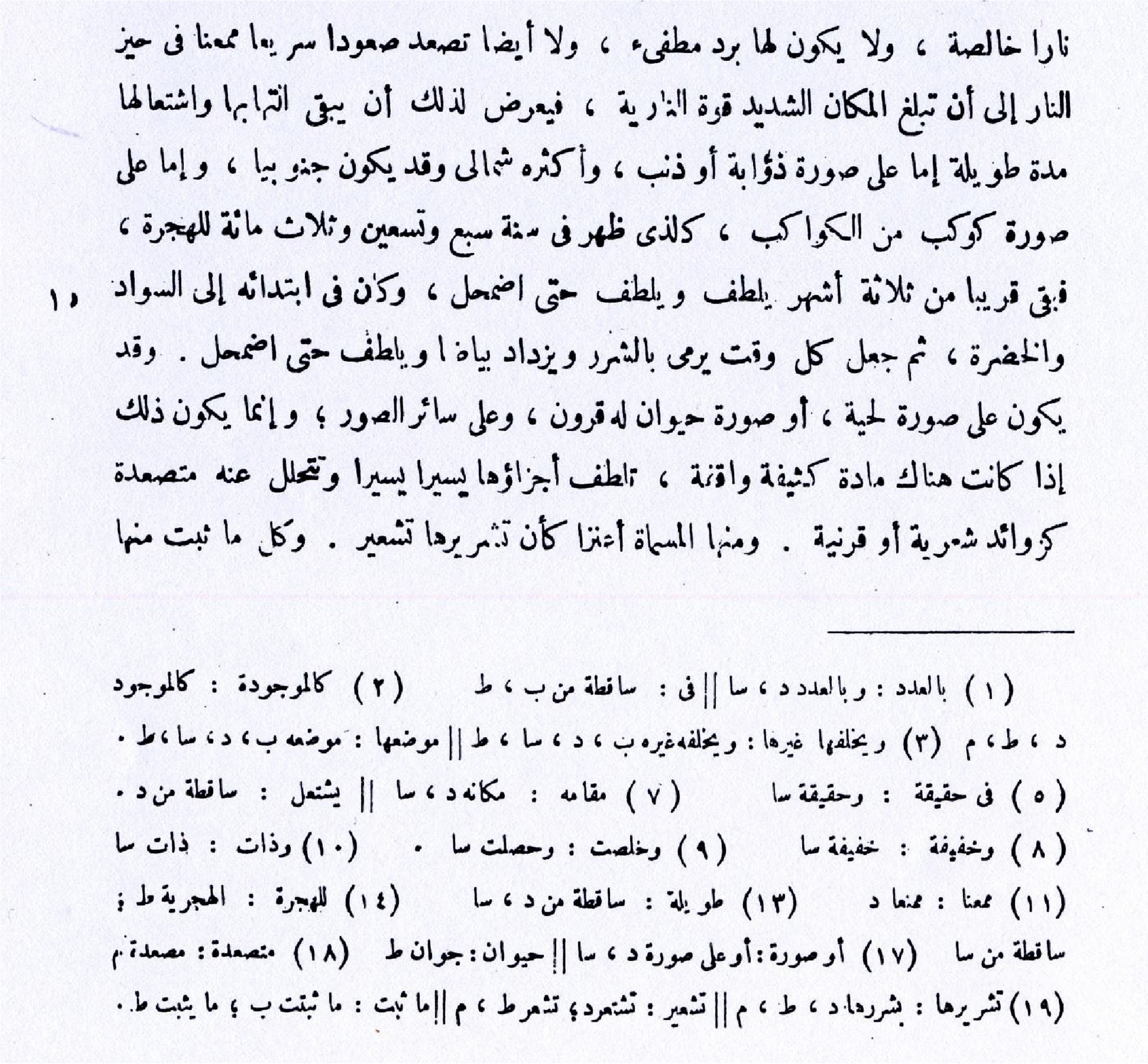}
\end{center}
\caption{Here we show the Arabic text from the report of SN 1006 of Ibn S{\={\i}}n\={a} in {\em al-Shif\={a}'}
from the Arabic edition by Madk\={u}r et al. (1965), page 73.
The relevant text starts in the middle of the second line from the top and ends almost at the (leftmost) end
of the 3rd-to-last line from the bottom of the main text.
The writing in the left margin is the Arabic line number 15.
The 4th line (line 14) reads (starting from the right) for the 2nd to 4th word {\em kawkab min al-kaw\={a}kib},
i.e. {\em a star among the stars}, and at the end of that line it specifies the year (the leftmost word is {\em hijra}).
The lines at the bottom indicate variant readings in different manuscripts, none of which change
the content and meaning of the relevant text about the new star:
the words for {\em long} and {\em hijra} are missing in one or two manuscripts.}
\end{figure*}

In the relevant chapter 5 of {\em al-Shif\={a}'}, Ibn S{\={\i}}n\={a} discussed the {\em Meteorology} of Aristotle.
Following Aristotle, Ibn S{\={\i}}n\={a} explains that most atmospheric optical phenomena 
and in particular those connected with humid air
would be due to wet {\em anathymiasis} (evaporation), while all phenomena connected with thunder, 
blizzard, wind as well as meteors
and comets (maybe including other transient celestial objects) would be due to dry {\em anathymiasis}. 
In the first sentence of the short quotation above, Ibn S{\={\i}}n\={a} obviously 
talks about what we today call comets ({\em form of a lock of hair or with a tail}).
After describing the transient star of 397h, he continues to talk about 
other transient objects including what we call comets ({\em it can also have the form of a beard ...}).
He says that such an object {\em stays for a (long) while}, i.e. that it is transient.
That the new {\em star among the stars} is discussed together with what we call comets 
is not surprising, as both refer to variable phenomena placed in the sub-lunar sphere.
In other words, the term {\em comet} was in former times used for several kinds
of transient objects including what we today call comets, novae, and supernovae.

Since Ibn S{\={\i}}n\={a} could have been an eyewitness of SN 1006,
let us consider where he was living, when SN 1006 was visible:
Ibn S{\={\i}}n\={a} left Bukhara (now Uzbekistan) between AD 999 and 1005 and went 
via Nishapur (Iran) at a geographic latitude of $36^{\circ}13^{\prime}$ north
and Merv (now Turkmenistan) at $37^{\circ}40^{\prime}$ north
to K\={a}th (now Uzbekistan) at $41^{\circ}41^{\prime}$ north, 
the ancient capital of the province of Khorasan south of the Aral lake,
which is now called Beruni in honor or
the Arabic scholar al-B{\={\i}r\={u}n{\={\i}, who was born here.
Ibn S{\={\i}}n\={a} left K\={a}th in AD 1012.
Hence, if Ibn S{\={\i}}n\={a} was an eyewitness of the new star in AD 1006/7,
he would have observed it most likely from a location as far north
as $36^{\circ}13^{\prime}$ to $41^{\circ}41^{\prime}$, probably the latter.
The text does not unfold whether Ibn S{\={\i}}n\={a} was the observer himself.

\section{Interpretation}

Let us first consider the dating of the observation, then the other information content.

\subsection{Dating}

An apparent difference to other reports is the year of appearance mentioned here by Ibn S{\={\i}}n\={a},
namely the year 397h, while the other Arabic reports all give 396h.

The Muslim calendar is a lunar calendar, where the months (and years) start with the evening when a
new crescent moon is seen (days run from evening to evening), see Quran, Sura 2, 189. There are no leap months.
Months can usually last 29 or 30 days, since the synodic month lasts 29.26 to 29.80 days (29.53 days on average).
If in history the Muslim months had alternating 29 or 30 days,
then, in any period of 30 years, there should be 11 years with one month with one extra day
(e.g. a lunar year with seven months with 30 days plus five months with 29 days).

The year is given in the Muslim Hijra era, i.e. the number of years
after the start of the lunar year in which the Hijra took place,
i.e. the emigration of the Islamic Prophet Mu\d{h}ammad from Mecca to Medina, known as Hijra.
This era, i.e. the year 1h started on AD 622 Jul 16/17 according to most scholars -- but it may have been on AD 622 Jul 15/16 according
to, e.g., de Blois (2000).

Any date given in the Muslim calendar can be converted to a Julian or Gregorian date with a precision of $\pm 2$ days.
The reason for this uncertainty is, among the uncertainty of the start of the era (see above),
that it is not clear a-posteriori when in history a month had an extra day\footnote{The artificially constructed 
{\em calculated} Islamic calendar uses leap days in certain, pre-defined years and months; 
in reality, we have to expect that, in each period of 30 years,
there were 11 months which had an additional day (due to real, late crescent sighting) --
in addition to those 354 days in a pure lunar calendar year with -- on
average -- six months of 29 days plus  -- on average -- six months of 30 days
(the average synodic month length is 29.53 days, and not 29.50).
Due to late crescent sighting, the month with one extra day did not necessarily follow the leap day/month rule in the
calculated Islamic calendar used in, e.g., Spuler \& Mayr (1961).
Hence, this calendar can deviate by up to two days (Ginzel 1906; Spuler \& Mayr 1961; de Blois 2000).}
and that the first sighting of a new crescent moon can be delayed due to, e.g.,
bad weather and/or difficult landscape.
It is also possible that -- even an experienced observer -- claims to have seen the crescent new moon,
even if it was not yet possible, so that a month would start one day too early (Doggett \& Schaefer 1994).
See, e.g., Spuler \& Mayr (1961), de Blois (2000), Said et al. (1989), 
and Neuh\"auser \& Kunitzsch (2014) for more details about Muslim calendar rules.

Since the date given by Ibn S{\={\i}}n\={a} is very rough, just the year is given,
we do not need to try to date the start of the year with high precision.
According to the {\em calculated} Islamic calendar (Spuler \& Mayr 1961),
the year 397h started on AD 1006 Sept 26 ($\pm 2$) in the evening.

As mentioned before, all other observers gave much earlier dates for their first observation:
$^{c}$Al\={\i} ibn Ri\d{d}w\={a}n observed SN 1006 since AD 1006 Apr 30,
the Chinese observers since May 1 (but possibly already on Apr 3), 
and in Japan, it was sighted first on Apr 28 or 30 (see Goldstein 1965; SG02);
it is possible that SN 1006 was already observed on around Apr 17/18 in Yemen,
see Rada \& Neuh\"auser (2015) for the evidence.
The observer in St. Gallen reports to have observed the new star {\em for three months} (SG02); 
if that was all after sunset, given his location, he cannot have observed it after about July 10, 
so that he probably started to observe SN 1006 in April 1006.
While most Arabic observers mention that they observed the SN for some 2-4 months,
the Moroccan report mentions six months as visibility period, 
which would be until after conjunction with the Sun (Goldstein 1965; SG02). 
The Chinese have observed also the helical setting and rising of SN 1006 in AD 1007 (SG02).
Hence, SN 1006 was still visible in the Muslim lunar year that started on AD 1006 Sept 26 ($\pm 2$)
in the evening.
SN 1006 was in conjunction with the Sun from mid Sept to mid Nov, so if it was observed in
397h, then it must have been after mid Nov 1006.

There are four possibilities to be considered for the interpretation of the year (397h) given by Ibn S{\={\i}}n\={a}:
\begin{enumerate}
\item Ibn S{\={\i}}n\={a} as eye-witness mistakenly gave the slightly wrong year in his text,
e.g. a memory error or typo, possibly years after the observation, 
{\em al-Shif\={a}'} was written AD 1014-1020.
\item The year was changed by mistake during the (oral ?) transmission from the 
observer to Ibn S{\={\i}}n\={a} from 396h to 397h.
\item A copying scribe made a mistake by changing the year from 396h to 397h.
\item The observer (possibly Ibn S{\={\i}}n\={a} himself) in fact observed SN 1006 in 397h, 
which started on AD 1006 Sept 26 ($\pm 2$), i.e. at or after heliacal rising after conjunction with the Sun
(after mid Sept).
\end{enumerate}
The latter (4) is less likely, because the star was brighter in the year 396h.
The year 397h most certainly is given by mistake -- just one year too late.

There is no other transient celestial object that 
could have been meant by Ibn S{\={\i}}n\={a} for 397h (AD 1006 or 1007),
in particular no comet (see Ho Peng Yoke 1962 and Kronk 1999).\footnote{Cook (1999) quoted some additional text 
from Ya\d{h}y\={a} ibn Sa$^{c}$\={\i}d al-An\d{t}\={a}k\={\i} after the
report for SN 1006 on what is most certainly a different transient object, 
e.g. a meteor or bolide: {\em Another star appeared
with a strong light in the west during the time of the falling of night during Saturday 
night, 9 Shawwal [10 July], and it stayed
long and grew great. Then it broke up into three parts and disappeared.} Appearance of the object in the {\em west}
at {\em falling of night} is not consistent with SN 1006, which was within an azimut 1h of the meridian at sunset. 
Even if the text here says {\em it stayed long and grew great},
this can be a bolide, which {\em stayed} relatively {\em long} for a bolide or meteor. The source for the latter object
(bolide) can very well be a different source than for the SN report. SG02 did not even include this extra text in their
citation of Ya\d{h}y\={a} ibn Sa$^{c}$\={\i}d al-An\d{t}\={a}k\={\i}, obviously also because they consider it unrelated to SN 1006. 
Alternatively, the last sentences, if pertaining to SN 1006, could either confirm the
long visibility period and strong brightness (until July) or might be mis-dated
(first appearance was April/May instead of July as given here).
The wording {\em it broke up into three parts and disappeared} might be understandable as
effect of strong scintillation at very low altitude, possibly around heliacal setting.
When the sun was some $9-18^{\circ}$ below the horizon, SN 1006 was clearly visible in the west 
(azimuth being 1.5 to 2.5h west of meridian).}
Further circumstantial evidence for a visibility period in spring or mid AD 1006 is given by the fact
that Ibn S{\={\i}}n\={a} specified that the object was visible for {\em close to three months},
which is quite consistent with the Arabic observers, who detected the object in April or May
and monitored it for typically two to four months.

\subsection{Stationarity, appearance, direction, duration, lightcurve, colour, and brightness}

We can now discuss the other information content from Ibn S{\={\i}}n\={a} and compare it to other observers.

{\bf Taillessness (and/or stationarity).} 
With the wording {\em in form of a star among the stars}, 
Ibn S{\={\i}}n\={a} probably means that the transient new object was tail-less --
in contrast to the more common transient objects, comets with tails ({\em if form of a lock of hair/beard}),
which move relative to the stars.
The wording {\em a star among the stars} may also or alternatively mean stationarity.
Other Arabic records mentioned the stationarity:
$^{c}$Al\={\i} ibn Ri\d{d}w\={a}n ({\em It remained where it was and it moved daily with its zodiacal sign}),
Ibn al-Jawz{\={\i}} wrote {\em ... and it remained fixed ...} (Goldstein 1965; SG02), 
and maybe also al-Yam\={a}n{\={\i}} and Ibn al-Dayba$^{c}$ ({\em remained unchanged}).

{\bf Direction.}
Even though SN 1006 indeed appeared in the far south as seen from Arabia or Persia for Ibn S{\={\i}}n\={a},
we cannot conclude on the direction from his text.
Even when he says that
[comets appear] {\em mostly in the north, but sometimes also in the south,
or in form of a star among the stars like the one which appeared in the year 397h},
he may just quote Aristotle for (normal) comets (or, more generally, transient celestial objects) 
to appear in both the north and the south,
before then starting to discuss the new star of AD 1006.

{\bf Duration.}
The duration of visibility given ({\em qar{\={\i}}ban min thal\={a}that ashhur} for {\em close to three months}) 
can mean a little less or a little more than three months,
and it is consistent with most other observers:
$^{c}$Al\={\i} ibn Ri\d{d}w\={a}n (four months), 
Ibn al-Jawz{\={\i}} and Ibn al-Ath\={\i}r ({\em beginning 
of Sha$^{c}$b\={a}n ... until the middle of Dh\={u} al-Qa$^{\rm c}$dah}, i.e. 3.5 months),
Moroccan report ({\em This star stayed for six months}),
Ya\d{h}y\={a} ibn Sa$^{c}$\={\i}d al-An\d{t}\={a}k\={\i} ({\em it continued four months}, Cook 1999),
al-Yam\={a}n{\={\i}} ({\em On the night of mid-Rajab in the year 396h, a star appeared ... 
In the night of mid-Rama\d{d}\={a}n, its light started to decrease and gradually faded away}, 
i.e. more than two months), and similarly
Ibn al-Dayba$^{c}$ ({\em on the night of mid-Rajab a star like Venus appeared ... It remained 
unchanged until the night of mid-Rama\d{d}\={a}n}, i.e. not less than two months).
In particular if Ibn S{\={\i}}n\={a} (or his source) observed SN 1006 since about late April or early May 1006,
and then for {\em close to three months}, then he could have observed until heliacal setting
for his location.

Given that Ibn S{\={\i}}n\={a} was located quite far north in AD 1006, see Sect. 2, 
(either in Nishapur $36^{\circ}13^{\prime}$ north or Merv $37^{\circ}40^{\prime}$ north,
or K\={a}th $41^{\circ}41^{\prime}$ north, probably the latter),
he could not observe SN 1006 for a long period (if he was the observer himself at all):
His likely location K\={a}th has a similar geographic latitude as Naples, Italy ($41^{\circ}$ north),
from where a new star is related to a 3 month period:
{\em A very brilliant star shone, and a large drought happened for three months}
(SG02 from the Annales Beneventani).
St. Gallen is even further north at $47^{\circ}25^{\prime}$, and surrounded by high mountains,
where the new star was also seen for 3 months (SG02).
As for the observer(s) in Naples,
Ibn S{\={\i}}n\={a} could have observed SN 1006 only until about the end of July
(if he observed only after sunset): On 1006 Jul 31, SN 1006 would have been at an
altitude of about $5^{\circ}30^{\prime}$ above the horizon at sunset with an apparent magnitude of about $-1$ mag.
(We also take into account that the southern horizon is quite flat as seen from K\={a}th/Beruni towards the south:
K\={a}th/Beruni is located at today's border of Uzbekistan to Turkmenistan and the latter has almost
no high mountains, in particular not south of K\={a}th/Beruni.)
Hence, the observer (whose report is transmitted by Ibn S{\={\i}}n\={a}) could have observed 
SN 1006 in May, June, and July (not later, but possibly earlier in April).
He may have observed (part of) the rise, the peak, the decrease, and even the heliacal setting of SN 1006.
The fact that Ibn S{\={\i}}n\={a} reports to have seen the new star for close to three months
(probably May, June, July) is fully consistent with the other Arabic reports.\footnote{For the observer 
in St. Gallen, the situation was even worse: he could observe SN 1006 only until July 10 at most given his even more 
northern location and the high mountains towards his southern horizon; since he observed for three months, 
he must have seen SN 1006 for his first time before May.}

{\bf Light curve.}
Ibn S{\={\i}}n\={a} also describes that the new star decreased in brightness before it disappeared
({\em getting fainter and fainter until it disappeared} and later
{\em and then became fainter and disappeared}), 
as do the Yemeni authors:
Al-Yam\={a}n{\={\i}} ({\em In the night of mid-Rama\d{d}\={a}n, its light started to decrease and gradually faded away})
and similarly Ibn al-Dayba$^{c}$ ({\em its light diminished and it gradually faded away}).
The connection of a gradual decrease in brightness with disappearance by Ibn S{\={\i}}n\={a} could well
mean the process of heliacal setting. Because SN 1006 was observed by some observers
even after conjunction with the Sun since the end of AD 1006, it is less likely that
SN 1006 was not observable any more due to intrinsic faintness before heliacal setting.  
The Arabic word {\em fa-baqiya} in the text by Ibn S{\={\i}}n\={a} 
was translated here as {\em It remained} (for close to three months),
which probably does not mean {\em It remained (fixed at its location)},
but rather in a temporal sense (that it was seen for close to three months),
or possibly that {\em it remained (somewhat constant in brightness)};
also, stationarity may have been already mentioned by Ibn S{\={\i}}n\={a} in his previous
sentence ({\em a star among the stars}).

{\bf Colour.} 
The colour and its evolution is mentioned:
{\em at the beginning it was towards a darkness and greenness,
then it began to throw out sparks [yarm{\={\i}} bi-l-sharar], and then it became more and more whitish ...}.
This part of the text may be more difficult to understand.
What we translated as {\em darkness} could even mean {\em blackness} or possibly {\em faintness}, 
but {\em black} is almost impossible as colour for a celestial object;
while the meaning of {\em black} as {\em unfortunate} was also known and
used in Arabic, this interpretation is unlikely here, because it is combined with {\em greenness} --
and also because Ibn S{\={\i}}n\={a} is known to have opposed {\em black magic}, astrology etc.
The wording {\em darkness and greenness} could mean {\em faint green-to-yellow} 
(at the beginning, i.e. before peak brightness).
What is described as {\em green} in the sky is often {\em yellow} or {\em yellowish-to-greenish},
as yellow and green are very similar to each other.
The Chinese have reported that the new star was {\em yellow} (SG02).
Ibn S{\={\i}}n\={a} continues with {\em then it became more and more whitish}.
In this case, Ibn S{\={\i}}n\={a} would report that the new star was first {\em yellowish-to-greenish}
and then {\em more and more whitish}.

While it is also not clear whether our translation of the colours shows the intended meaning
or even whether the transmitted text is somewhat corrupt in this part,
his text can be interpreted in a consistent way as follows: \\
{\em At the beginning it was towards a faintness and greenness}
meaning that it was faint and greenish-yellowish at the beginning (like the Chinese report: 
it {\em increased in brightness} and was {\em yellow}, see SG02),
{\em then it began to throw out sparks all the time, and then it became more and more whitish},
i.e. that it was scintillating during the period of largest brightness (similar in China
and in other Arabic reports) being more and more whitish (brighter?) during peak brightness.
The colour evolution reported is independent from other Arabic reports, 
both regarding the content and details and regarding the Arabic wording.

{\bf Brightness.}
Ibn S{\={\i}}n\={a} reports that it {\em throws out sparks}, i.e. that it scintillated, 
this is again quite consistent with the other Arabic reports, consistent with a very bright luminosity:
$^{c}$Al\={\i} ibn Ri\d{d}w\={a}n ({\em it twinkled very much}),
Ibn al-Jawz{\={\i}} ({\em It was glittering}),
Ibn Sa$^{c}$\={\i}d al-An\d{t}\={a}k\={\i} ({\em It had dazzling rays and a great rippling}),
as well as al-Yam\={a}n{\={\i}} and Ibn al-Dayba$^{c}$ ({\em It showed a great turbulence}).
The wording {\em yarm{\={\i}} bi-l-sharar} for {\em throw out sparks} for scintillation is
different from other Arabic reports, again showing its independence.
Since Ibn S{\={\i}}n\={a} does not compared the new star with either the brightest star(s) in the sky 
nor with Venus (the brightest planet in the sky), the new star was probably much brighter than
the brightest stars -- and even much brighter than Venus, i.e. brighter than about $-5$ mag,
the largest possible brightness of Venus.

\section{Summary}

The presented record by Ibn S{\={\i}}n\={a} about a transient celestial object in 397h 
is quite clearly related to other credible reports about SN 1006, but original
(however, dated one year too late).
Within a discussion about presumably sub-lunar phenomena, Ibn S{\={\i}}n\={a} reports 
a transient celestial object in form of a tail-less {\em star among the stars}, 
which was seen for close to three months, and it was scintillating;
its colour may have changed first from greenish(-yellowish) to whitish, 
and then it gradually decreased in brightness (probably due to heliacal setting).

In general, Ibn S{\={\i}}n\={a}'s text is consistent with other Arabic (and non-Arabic) records about SN 1006,
the main addition is the colour evolution and some terms, e.g. {\em star among the stars}
and {\em yarm{\={\i}} bi-l-sharar} for {\em throw out sparks}.
Hence, the report is independent.

When $^{c}$Al\={\i} ibn Ri\d{d}w\={a}n told us that {\em other scholars from time to time have followed it
[SN 1006] and came to a similar conclusion} (SG02), he may have meant Ibn S{\={\i}}n\={a}, among others.

\acknowledgements
We thank the Institut f\"ur Geschichte der Arabisch-Islamischen Wissenschaften, Frankfurt,
where we could use the extensive library with Ibn S{\={\i}}n\={a}'s {\em al-Shif\={a}'} in Arabic.
We also acknowledge Dagmar L. Neuh\"auser for various important comments.

{}

\end{document}